\documentclass[5p,twocolumn,times,number]{elsarticle}

\usepackage{graphicx}
\usepackage{amsmath}
\usepackage{multirow}

\begin{document}

\begin{frontmatter}

\title{Digital signal processing for a thermal neutron detector using ZnS(Ag):${}^6$LiF scintillating layers read out with WLS fibers and SiPMs}

\author{J.-B.~Mosset\corref{cor}}
\ead{jean-baptiste.mosset@psi.ch}
\author{A.~Stoykov}
\author{U.~Greuter}
\author{M.~Hildebrandt}
\author{N.~Schlumpf}
\cortext[cor]{Corresponding author}
\address{Paul Scherrer Institut, CH-5232 Villigen PSI, Switzerland}

\begin{abstract}
We present a digital signal processing system based on a photon counting approach which we developed for a thermal neutron detector consisting of ZnS(Ag):${}^6$LiF scintillating layers read out with WLS fibers and SiPMs. Three digital filters have been evaluated: a moving sum, a moving sum after differentiation and a digital CR-RC${}^4$ filter. The performances of the detector with these filters are presented. A full analog signal processing using a CR-RC${}^4$ filter has been emulated digitally. The detector performance obtained with this analog approach is compared with the one obtained with the best performing digital approach.



\end{abstract}


\end{frontmatter}

\section{Introduction}
A 1-D position sensitive detector for thermal neutrons is currently under development \cite{Hildebrandt,Stoykov_2015,Mosset_2014b} for upgrading the POLDI instrument, a strain-scanning diffractometer installed at the Swiss neutron spallation source (SINQ) at PSI. This detector is based on ZnS(Ag):${}^6$LiF scintillating layers read out with wavelength shifting fibers (WLS) and silicon photomultipliers (SiPM).

The main challenges concerning the signal processing are the reduction of the detector background count rate due to the SiPM dark counts down to $10^{-3}$~Hz and the reduction of the multi-count ratio due to the scintillator afterglow down to $10^{-3}$, together with a high trigger efficiency and a high neutron count rate capability.

\section{Principle of the signal processing system (SPS)}

\begin{figure}[h!]
\centering
\includegraphics[width=1\linewidth]{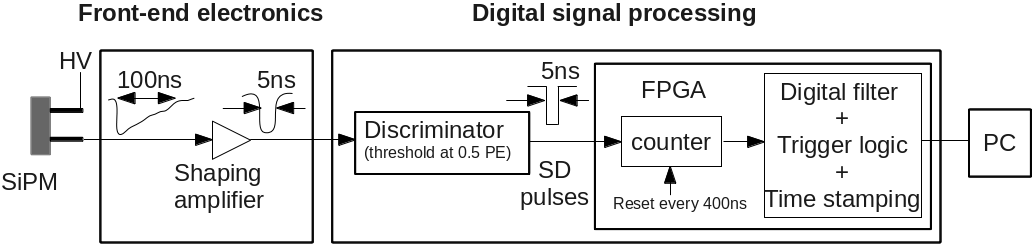}
\caption{Block diagram of the signal processing system.}
\label{fig:block_diagram}
\end{figure}

Figure \ref{fig:block_diagram} shows the block diagram of the SPS. The SiPM signal is amplified and shaped by a wide band-width amplifier. The output of the amplifier is then fed into a fast leading edge discriminator with a threshold set at 0.5 photoelectron. During each consecutive time slice $\Delta t$ of 400~ns (for example), the number of SD pulses after the discriminator is measured. In other words, the density of SD pulses in time is sampled. 

Once sampled, the signal is filtered and the following triggering conditions are ANDed:
a) The channel is ready (after each trigger, an artificial dead time is introduced to prevent multiple triggers on the same events).
b) The filter output is maximum.
c) The filter output is higher than a certain threshold.

The time stamp of an event is defined by the time slice during which the filter output is maximum.

\section{Digital filters under evaluation}

This section presents the discrete-time implementation of the three digital filters which have been evaluated. We denote $x_i$ and $z_i$ the discrete samples of the signal at the input and output of the filter, respectively.\vspace{1.5mm}\\
a) Moving sum (MS) of span $M$ \hfill $z_i = z_{i-1} + x_{i} - x_{i-M}$\vspace{1.5mm}\\
{\raggedright b) Moving sum after differentiation (DI+MS)\\}
$y_i = x_{i} - x_{i-M}$ (DI) \hfill $z_i = z_{i-1} + y_{i} - y_{i-M}$\vspace{1.5mm}\\
c) CR-RC${}^4$ \hfill $z_i = \sum_{j=1}^{5} b_j z_{i-j} + \sum_{j=1}^{4} a_j x_{i-j}$\\
where $a_j$ and $b_j$ are non-integer constants involving the time constant $RC$ and the sampling interval $\Delta t$ \cite{Nakhostin_2011}.

\section{Evaluation of the digital filters}

The evaluation is based on real data which provide for each detected neutron the temporal sequence of SD pulses produced during the first 80~$\mu$s following the neutron capture. The data file contains 10~kevents. During the measurement, the rate of captured neutrons was 20~Hz and the SiPM dark count rate was 64~kHz. In order to evaluate the filter performances at different SiPM dark count rates, temporal sequences of dark counts are simulated and merged to the measured data. SiPM crosstalk and afterpulses, as well as the saturation of the SD pulse counting due to the 12~ns dead time of the discriminator are implemented in the simulation.

\begin{figure}[h!]
\centering
\includegraphics[width=1\linewidth]{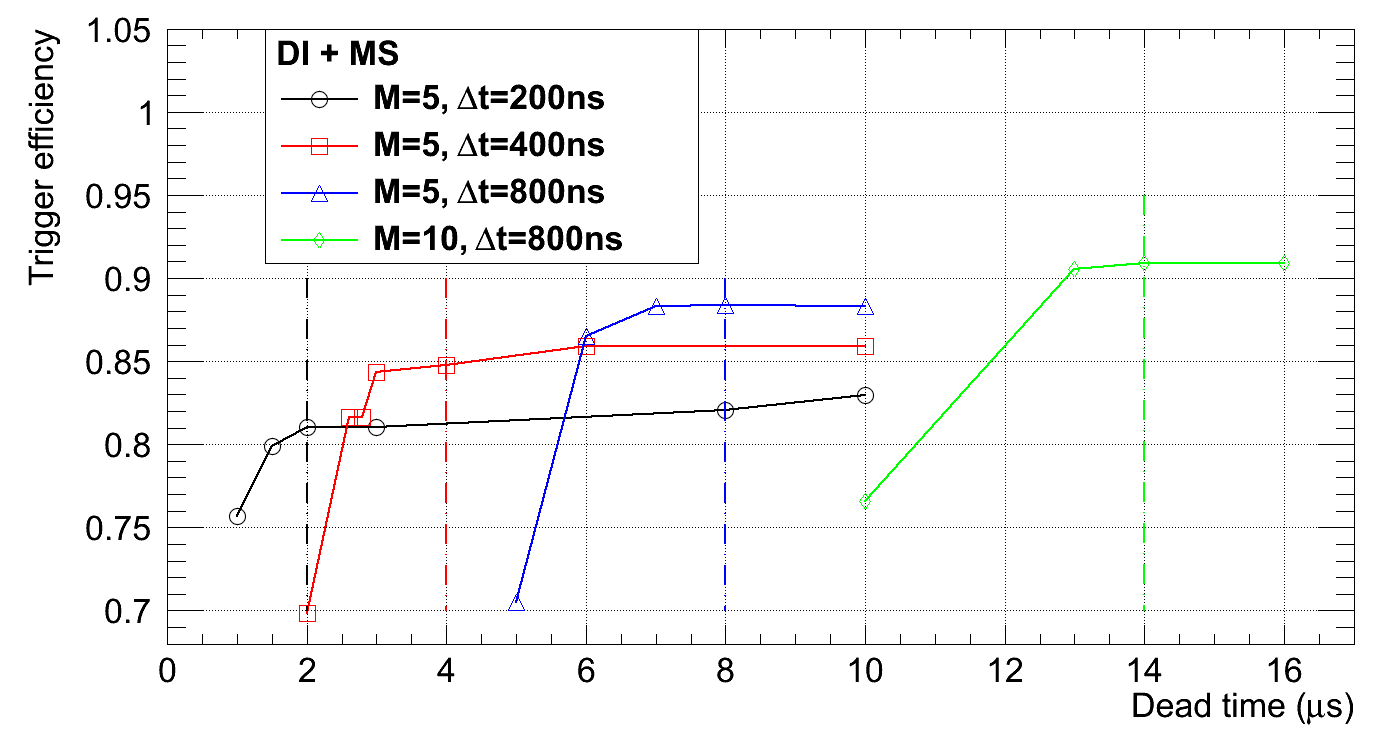}
\vspace{-4mm}
\caption{Trigger efficiency versus dead time for the DI+MS filter. The SiPM dark count rate is 63~kHz and the trigger threshold is adjusted to ensure a multi-count ratio $<10^{-3}$. The vertical dashed lines denote the minimum dead times which can be set without significant decrease of the trigger efficiency.}
\label{fig:dead_time_DI_MS_PW}
\end{figure}

\begin{figure}[h!]
\centering
\includegraphics[width=1\linewidth]{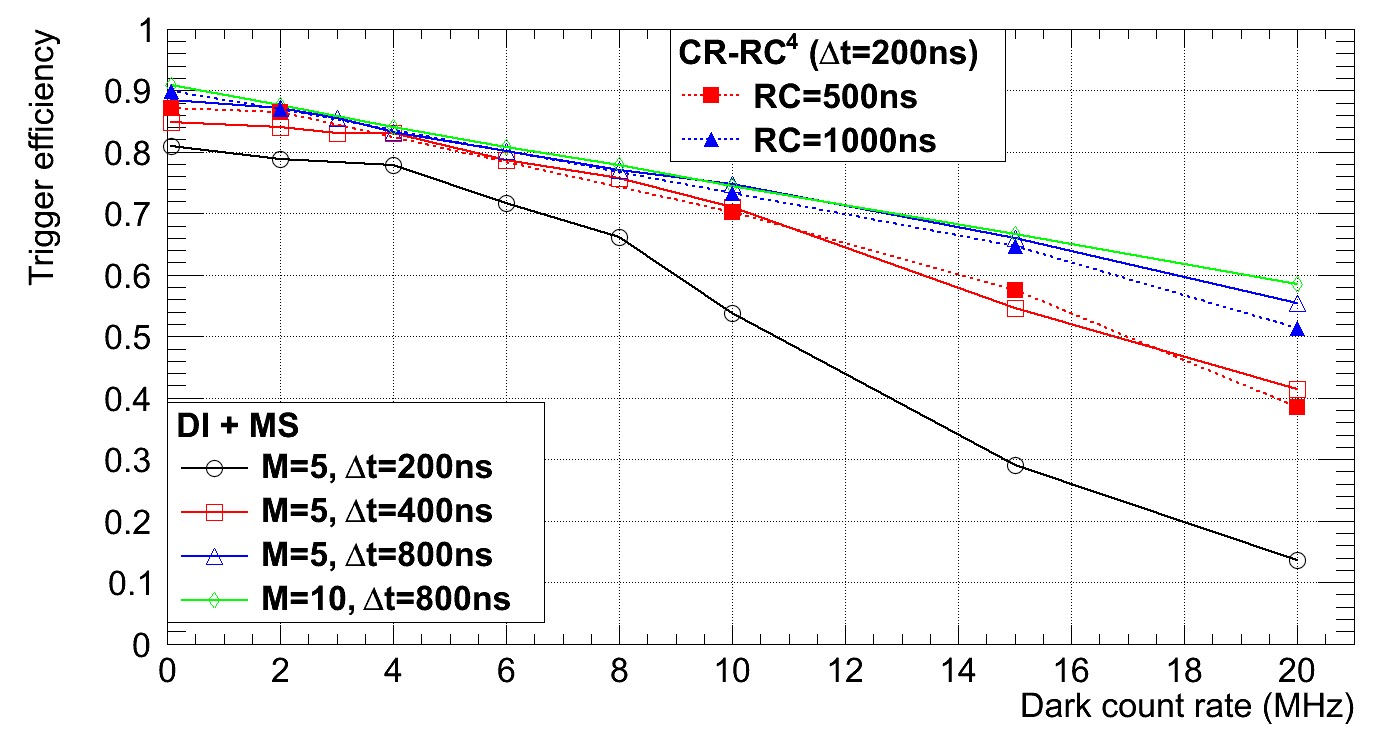}
\vspace{-4mm}
\caption{Trigger efficiency versus SiPM dark count rate. Conditions: background count rate $<10^{-3}$~Hz, multi-count ratio $<10^{-3}$, dead times given by table \ref{table:dead_time}.}
\label{fig:trig_eff}
\end{figure}

Figure \ref{fig:dead_time_DI_MS_PW} shows the trigger efficiency as a function of the artificial dead time for the DI+MS filter. The SiPM dark count rate is 63~kHz and the trigger threshold is adjusted for each value of the dead time so that the multi-count ratio is $<10^{-3}$. For each filter, we determine the minimum dead time which can be set without significant decrease of the trigger efficiency. Table \ref{table:dead_time} gives these minimum dead times $\tau$ for all the evaluated filters as well as the time resolutions $R_t$ (stdev) obtained under the following conditions: SiPM dark count rate of 2~MHz, background count rate $<10^{-3}$~Hz, multi-count ratio $<10^{-3}$. The time resolution is approximatitely constant up to a dark count rate of $\sim$~20~MHz. Figure \ref{fig:trig_eff} shows the trigger efficiency as a function of the SiPM dark count rate for the DI+MS and CR-RC${}^4$ filters, under the conditions that the background count rate and the multi-count ratio are below $10^{-3}$~Hz and $10^{-3}$, respectively.

\begin{table}[h]
\vspace{-5pt}
\caption{Dead times $\tau$ set for all the evaluated filters and time resolutions $R_t$ under the following conditions: SiPM dark count rate of 2~MHz, background count rate $<10^{-3}$~Hz, multi-count ratio $<10^{-3}$.}
\vspace{-2pt}
\begin{small}
\begin{center}
\begin{tabular}{l|c|c|c}
\hline
Filter & $M$ or $RC$ ($\mu$s), $\Delta t$ (ns) & $\tau$ ($\mu$s) & $R_t$ (ns)\\
\hline
\multirow{2}{*}{MS} & 5, 200 & 20 & 120\\
& 10, 200 & 40 & 210\\
\hline
\multirow{2}{*}{CR-RC${}^4$} & 0.5, 200 & 5 & 220\\
& 1, 200 & 9 & 360\\
\hline
\multirow{4}{*}{MS+DI} & 5, 200 & 2 & 75\\
& 5, 400 & 4 & 140\\
& 5, 800 & 8 & 320\\
& 10, 800 & 14 & 650\\
\hline
\end{tabular}
\end{center}
\label{table:dead_time}
\end{small}
\vspace{-9pt}
\end{table}

\begin{figure}[h!]
\centering
\includegraphics[width=1\linewidth]{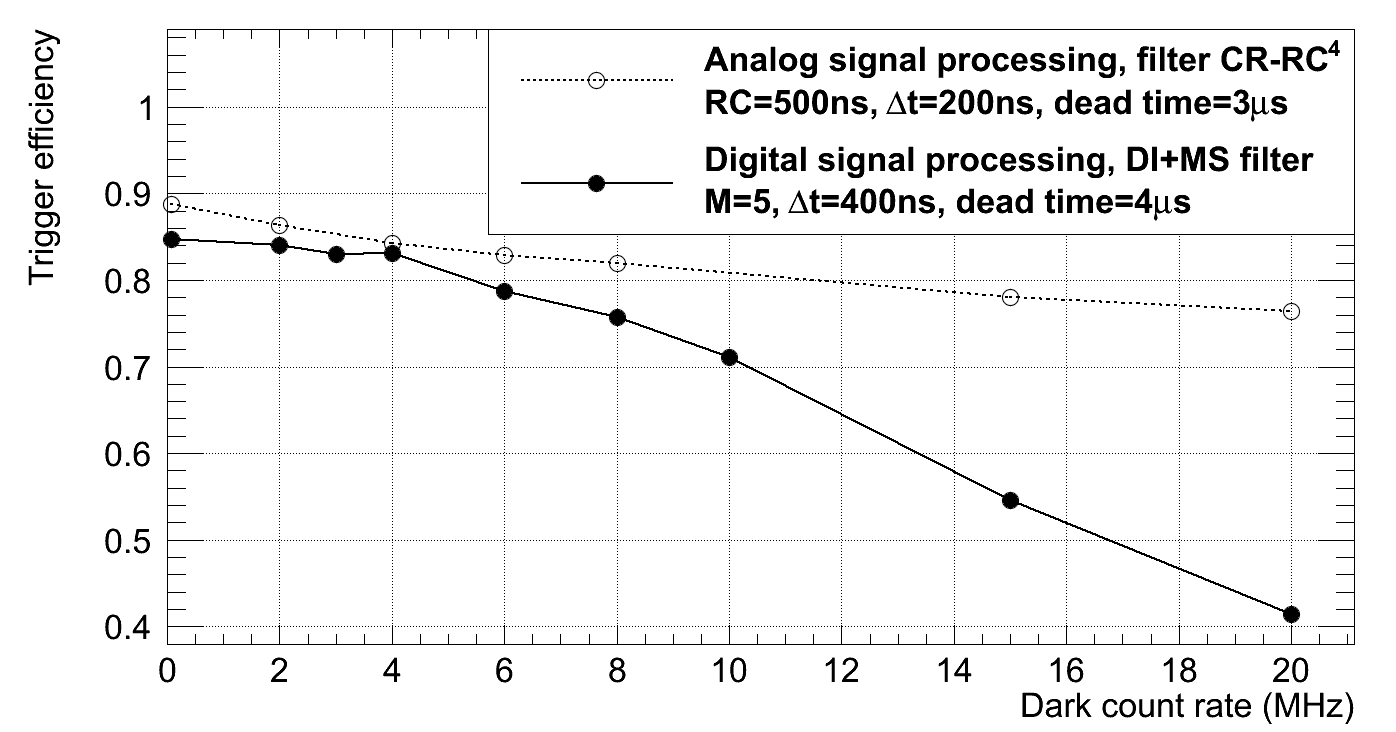}
\vspace{-4mm}
\caption{Trigger efficiency versus SiPM dark count rate for a digital approach using the DI+MS filter and for a full analog approach using a CR-RC${}^4$ filter. Conditions: background count rate $<10^{-3}$~Hz, multi-count ratio $<10^{-3}$.}
\label{fig:analog}
\end{figure}

To estimate the performances of an analog pulse processing, the measured data are corrected for the 12~ns dead time of the discriminator and the effect of the SiPM crosstalk is added. The discrete-time implementation of the CR-RC${}^4$ filter is then used to emulate an analog readout with a CR-RC${}^4$ filter. Figure \ref{fig:analog} shows the trigger efficiency as a function of the SiPM dark count rate for the digital approach using the DI+MS filter and for a full analog approach using a CR-RC${}^4$ filter.

\section{Conclusion}

The MS filter requires a significantly longer dead time than the other filters to ensure a multi-count ratio below $10^{-3}$. The DI+MS filter shows the same performance as the digital CR-RC$^4$ filter and it requires much less computational ressources. We report the following performance parameters of the detector with the DI+MS filter ($M$=5, $\Delta t$=400~ns): trigger efficiency $\geq$~80\%, dead time of 4~$\mu$s, time resolution (stdev) $< 170$~ns (contribution from the SPS), background count rate $< 10^{-3}$~Hz, multi-count ratio $< 10^{-3}$. All these parameters are achieved up to a SiPM dark count rate of 5~MHz. On the basis of preliminary irradiation measurements, we estimate that an increase of the SiPM dark count rate from the nominal value of 100~kHz to 5~MHz corresponds to more than 20~years of operation.

A full analog signal processing using a CR-RC${}^4$ filter has been emulated digitally. Up to 5~MHz SiPM dark count rate, the digital signal processing using the DI+MS filter has about the same performance as the full analog signal processing.

\end{document}